\documentclass{JINST}
\usepackage{graphicx}
\usepackage{fontenc}
\usepackage{times}
\usepackage{mathptmx}
\usepackage{amsfonts}
\usepackage{amsmath}
\usepackage{mathrsfs}
\usepackage{amssymb}
\usepackage{lineno}

\title{
Characterization of a broad energy germanium detector and
application to neutrinoless double beta decay search in $^{76}\mbox{Ge}$ 
}

\author{M.~Agostini$^{a,b,c}$, E.~Bellotti$^{d}$, R.~Brugnera$^{a,b}$, 
C.~M.~Cattadori$^{d}$, A.~D'Andragora$^e$, 
A.~di~Vacri$^e$, A.~Garfagnini$^{a,b}$, 
M.~Laubenstein$^e$, L.~Pandola$^e$\thanks{Corresponding author}~ 
~and C.~A.~Ur$^b$\\
\llap{$^a$} Dipartimento di Fisica \lq\lq G. Galilei\rq\rq\ , 
Universit\`a di Padova, \\ Via Marzolo 8, I-35131 Padova, Italy\\
\llap{$^b$} INFN Padova, \\ Via Marzolo 8, I-35131 Padova, Italy\\
\llap{$^c$} 
Physikdepartment E15, Technischen Universit\"at M\"unchen, \\
James-Franck-Strasse 1, D-85748 Garching, Germany\\
\llap{$^d$} INFN Milano Bicocca, \\
Piazza della Scienza 3, I-20126 Milano, Italy\\
\llap{$^e$} INFN, Laboratori Nazionali del Gran Sasso, \\
S.S. 17 bis km 18+910, 
I-67100 Assergi, L'Aquila, Italy\\

E-mails: \email{pandola@lngs.infn.it}}

\abstract{The performance of a 630~g commercial \emph{broad energy germanium} 
(BEGe) detector has been 
systematically investigated. Energy resolution, linearity, stability vs. 
high-voltage (HV) bias, thickness and uniformity of dead layers have been 
measured and found to be excellent. \\
Special attention has been dedicated to the study of the detector response 
as a function of bias HV. The nominal depletion voltage being 3000~V, the 
detector under investigation shows a peculiar behavior for biases around 2000~V: in a 
narrow range of about 100~V the charge collection is strongly 
reduced. The detector seems to be composed by two parts: a small volume 
around the HV contact where charges 
are efficiently collected as at higher voltage, and a large volume where 
charges are poorly collected. A 
qualitative explanation of this behavior is presented. An event-by-event pulse 
shape analysis based on $A$/$E$ (maximum amplitude of the current pulse  
over the total energy released in the detector) has been applied to events in different 
energy regions and found very effective in rejecting non localized events. 
In conclusion, BEGe detectors are excellent candidates for the 
second phase of GERDA, an experiment devoted to neutrinoless double 
beta decay of $^{76}$Ge.}


\begin{document}
\section{Introduction} \label{sec:intro}
This paper deals with the characterization of a 630~g \emph{broad energy germanium}
(BEGe) detector~\cite{bege}, a commercial high-purity Ge detector with a small
read-out electrode which is manufactured by Canberra
Semiconductor~\cite{canberra}. The work is part of
the general effort of the GERDA collaboration to improve the understanding of
these detectors which will be employed in the second phase of the experiment.
Preliminary results obtained with the detector under investigation  
have been already presented in Refs.~\cite{div09,ago09}.
A detector with the same contact geometry, differing only for dimensions, has
been previously investigated in the framework of the GERDA
activities~\cite{bu09,bu09b,ba10}. \\
The GERDA (GERmanium Detector Array) 
experiment~\cite{ge04}, which recently started its commissioning phase at the
underground Gran Sasso Laboratory of the INFN (Italy),  is devoted to the study of
the neutrinoless double beta decay of $^{76}$Ge with a
large mass of detectors isotopically
enriched in $^{76}$Ge; about 40~kg are planned for the final configuration. 
In searching for rare processes, like neutrinoless 
double beta decay, two parameters
are extremely important: energy resolution and background.  Substantial background
reduction can be achieved by a careful selection of materials used in the
set-up construction (not considered in this paper) and by an efficient data
analysis which needs in turn a deep knowledge of the detector
characteristics.\\
The characterization of the BEGe detector was divided into two parts:
experimental investigation of the detector features; modeling of the detector
signal formation and development.
The present note describes the experimental part while the detector modeling is
presented and discussed in Ref.~\cite{ago10} (companion paper).\\
The paper is organized as follows: Sect.~\ref{section:two} summarizes the
detector specifications provided by the manufacturer and the performance in
terms of active volume, linearity and energy resolution.  In
Sect.~\ref{section:three} the detector response to radiation sources
at different bias high voltages is described and discussed.  A peculiar behavior
of some BEGe detectors, already reported in Refs.~\cite{div09,bu09c,co10}, 
is reproduced experimentally and discussed in details.  
Sect.~\ref{section:four} reports the performance of the detector in
the discrimination of single-site to multi-site events by the analysis of 
the shape of the pulse signal. Finally, outlook and conclusions are presented in
Sect.~\ref{section:five}.\\
\section{Detector specifications and performance} \label{section:two}
\begin{figure}[tp]
   \begin{center}
      \includegraphics{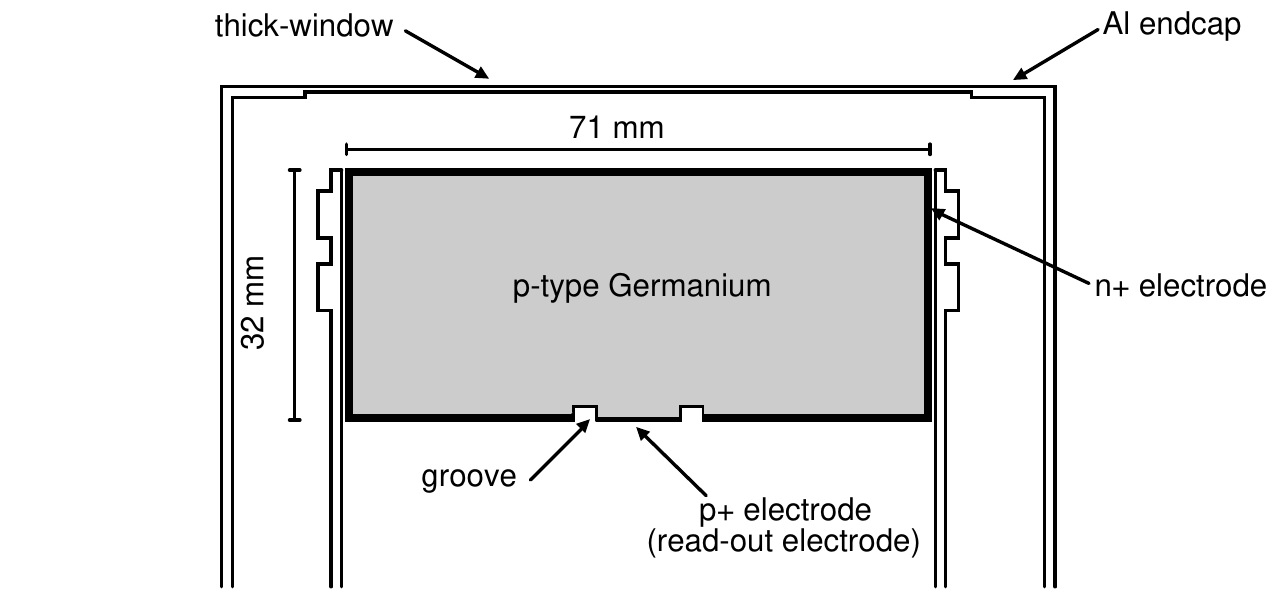}
   \end{center}
   \caption{Sketch of a BEGe detector mounted in its copper holder inside an
   Al cryostat. The signal read-out electrode and the groove are not to
   scale. The plot shows a vertical section of the detector passing through the
   symmetry axis.}
   \label{fig:detgeometry} 
\end{figure}

The detector under study is a commercial p-type \emph{broad energy germanium} (BEGe)
detector (model BE3830/s) manufactured by Canberra Semiconductor.  The detector
has a cylindrical shape with a diameter of 71~mm and a thickness of 32~mm. There
is a small Boron-implanted $p^+$ electrode on one of the flat surfaces. The
Lithium-diffused $n^+$ electrode covers the rest of the outer surface and it is
separated from the $p^+$ electrode by a circular groove. The detector is mounted
in a 1.5~mm thick cylindrical aluminum housing. A schematic view of the BEGe
detector and of its housing is presented in Fig.~\ref{fig:detgeometry}. 
The positive bias voltage is applied to the $n^+$ electrode while the $p^+$
electrode (signal read-out electrode) is grounded.  The detector specifications
and performance according to the manufacturer data sheet are summarized in
Tab.~\ref{tab::parameters}.  The bias voltage recommended by the manufacturer
and used in most the measurements is 3500~V.
%
%
\begin{table}[tp]
\caption{Geometrical and electrical parameters of the BE3830/s BEGe detector
produced by Canberra Semiconductor according to the manufacturer data sheet.
The nominal energy resolution at the 1~332.5~keV $\gamma$-line of $^{60}$Co is also 
reported. The resolution is quoted as full width at half of the
peak maximum - FWHM - and full width at one-tenth of the peak maximum - FW(1/10)M.
\label{tab::parameters}}
\begin{center}
\begin{tabular}{|l c|}
\hline
\multicolumn{2}{|l|}{Physical characteristics:}  \\
Active diameter & 71~mm \\
Active area & 3800~mm$^{2}$ \\
Thickness (active) & 32~mm \\
Distance from entrance window & 5~mm\\
Window thickness (Al) & 1.5~mm \\
Dead layer thickness & 0.8~mm\\
\hline
\multicolumn{2}{|l|}{Electrical characteristics:} \\
Depletion voltage & $+ 3000$~V$_{dc}$ \\
Recommended bias voltage & $+ 3500$~V$_{dc}$ \\
\hline
\multicolumn{2}{|l|}{Energy resolution at 1.332~MeV:}\\
FWHM & 1.752~keV (4 $\mu$s shaping time) \\
FW(1/10)M & 3.259~keV (4 $\mu$s shaping time) \\
\hline
\end{tabular}
\end{center}

\end{table}
%
%
\subsection{Data acquisition systems} \label{sec:elec}
The charge signal coming from the small $p^+$ read-out electrode  is sent to
the charge-sensitive pre-amplifier model 2002CS integrated in the detector
housing.  The pre-amplifier output signals are processed using two different
Data Acquisition (DAQ) systems, according to the aims of the measurement
considered:
\begin{enumerate}
\item A standard analogue electronic chain, with a spectroscopy amplifier Ortec
   672 and a multichannel Ethernim ADC. The data are then analyzed using the
   software MAESTRO-32 or GammaVision-32 provided by Ortec~\cite{ortec}.
\item A digital chain, in which the pre-amplifier output signals are fed to a
   CAEN 4-channel digitizer (module N1728B), with sampling frequency of 100~MHz 
   and 14-bit resolution~\cite{caen1728}.
   The digitizer is controlled by a computer connected via USB and running the
   TNT-TUC program distributed by CAEN~\cite{tuc}. 
   The digitized signals are transferred to
   the computer and analyzed by a set of \emph{ad hoc} programs~\cite{ge11}.
   Traces are acquired for 40~$\mu$s, including a 10~$\mu$s baseline before the 
   signal.
\end{enumerate}
%
%
%
%
%
%
\subsection{Linearity}
\begin{figure}[tbp]
\centering
\includegraphics[width=12cm]{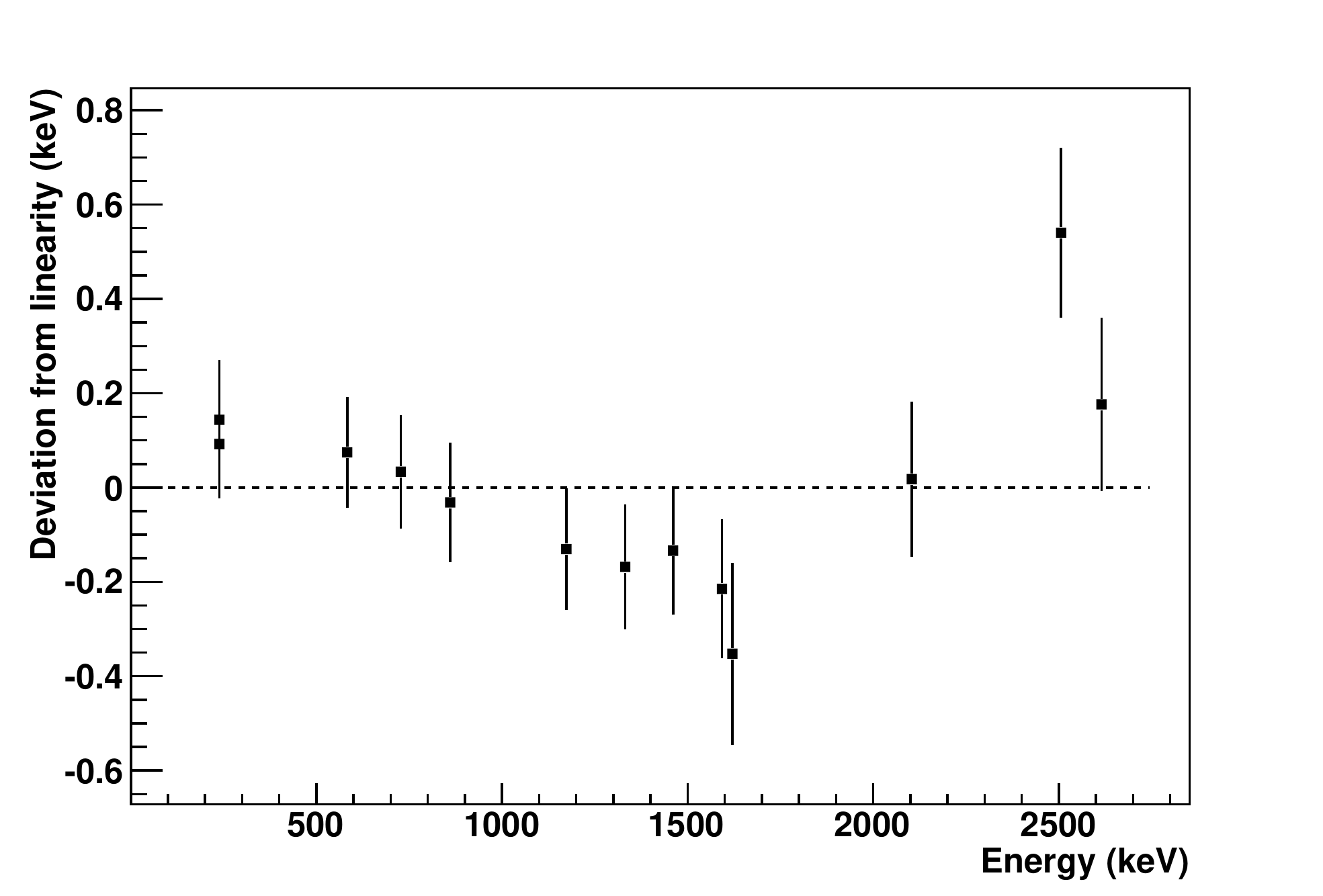}
\caption{Deviation from linearity of the BEGe detector response to different
$\gamma$-lines measured by the analogue DAQ system.  The two close
data points on the left are from the 238.6~keV and 241.9~keV $\gamma$-lines of
$^{212}$Pb and $^{214}$Pb, respectively.}
\label{fig:linearity}
\end{figure}

First, the linearity of the analogue electronic chain has been tested with an Ortec
Research Pulser module 448. It results to be better than 0.01\%.  Then the global
linearity of the system (i.e. detector and analogue electronic chain) has been
verified irradiating the crystal with $^{60}$Co and $^{228}$Th sources.  Deviations from
linearity were found to be less than 0.02\% in the energy range from 239~keV
up to 2.614~MeV.  Fig.~\ref{fig:linearity} shows the residuals of a linear fit,
i.e. the differences between data and the values expected from the  fit. 
The linearity obtained with the digital DAQ system and the off-line 
energy reconstruction is slightly worse, but in any case better 
than 0.05\%.
%
%
\subsection{Energy resolution}
\begin{figure}[tbp]
\begin{center}
\includegraphics[width=12cm]{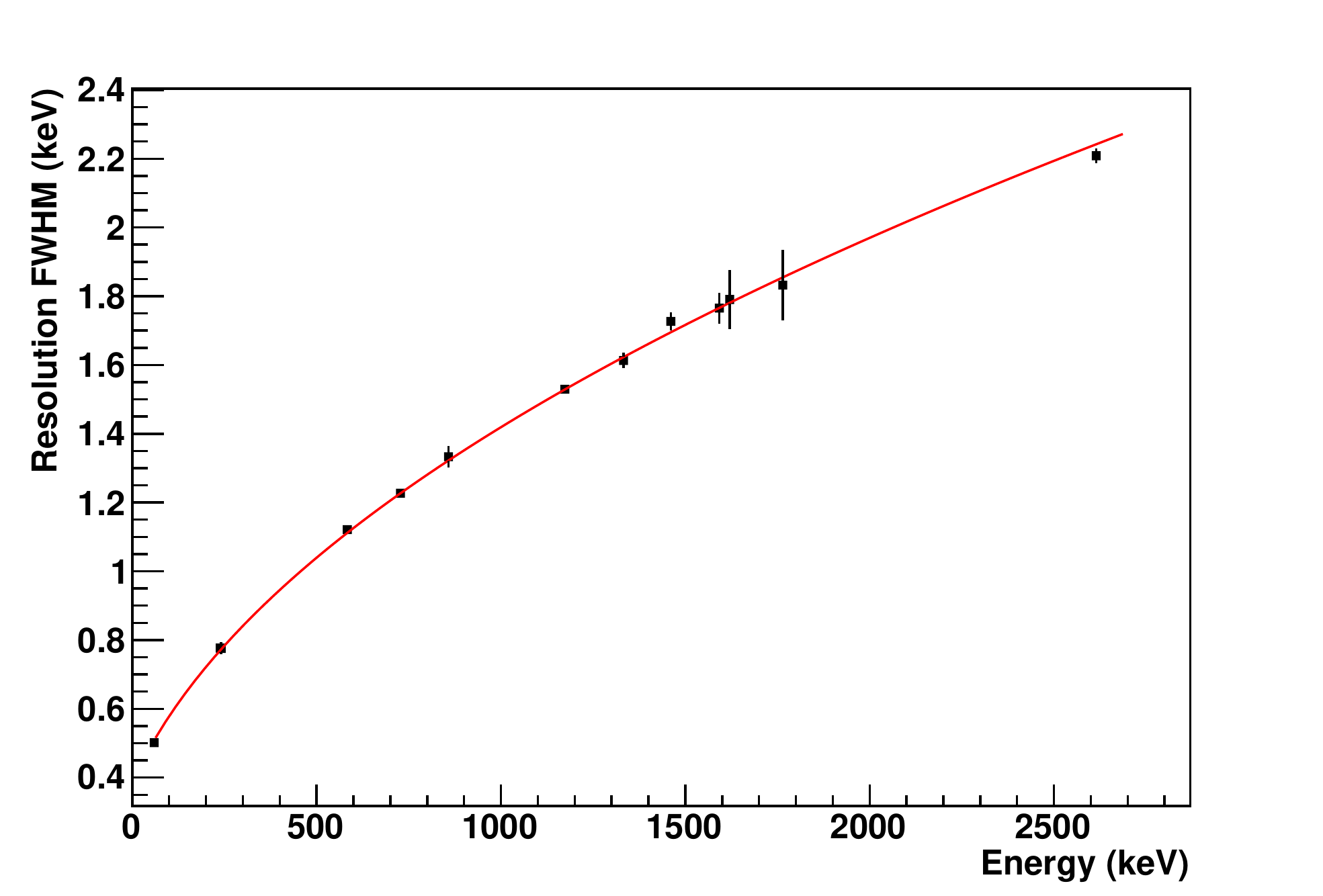}
\caption{Energy resolution (FWHM) as a function of energy for the analogue DAQ
system. 
The curve superimposed to the data points is the $\sqrt{a^2 + b E}$ parametrization.}
 \label{fig:resolution} 
\end{center}
\end{figure}

The energy resolution of the detector was studied by irradiating the crystal
with a $^{228}$Th and a $^{60}$Co source. 
The resolution obtained with the analogue DAQ system at the $^{60}$Co lines 
is compiled in Tab.~\ref{tab:resolution} for different shaping times.
The energy resolution achieved for 6~$\mu$s shaping time at the 1332~keV
$^{60}$Co line is $1.56 \pm 0.02$~keV FWHM, which is better than the manufacturer
specification. \\
Fig.~\ref{fig:resolution} shows the
energy resolution as a function of the energy for the $^{228}$Th and $^{60}$Co 
peaks in the spectra recorded by the analogue DAQ system.
The behavior is well described by the parametrization
\begin{equation}
   \mbox{FWHM} = \sqrt{a^2 + b E} \label{eq:resolution}
\end{equation}
with $a = (0.38 \pm 0.01)  \ \textrm{keV}$ and \
$b = (1.87 \pm 0.02) \cdot 10^{-3} \ \textrm{keV}$; the fit has $\chi^2=5.9$ with 
11 d.o.f.
\begin{table}[tp]
\caption{Energy resolution at the $^{60}$Co $\gamma$-lines 
for the analogue DAQ system with different values of the
shaping time.  The resolution is estimated as FWHM,
full with at one-fifth of the peak maximum  - FW(1/5)M - and FW(1/10)M.  The uncertainty
on the resolution - as evaluated from repeated measurements - is 0.02~keV.
Peaks have been modeled with a Gaussian curve sitting on a linear background.}
\label{tab:resolution}
\begin{center}
\begin{tabular}{|l | l | c c|}
\hline
Shaping time &  & 1173~keV & 1332~keV \\
\hline
3~$\mu$s & FWHM & 1.74 & 1.82 \\
 & FW(1/5)M & 2.62 & 2.84 \\
 & FW(1/10)M & 3.23 & 3.53 \\
\hline
6~$\mu$s & FWHM & 1.53 & 1.56 \\
 & FW(1/5)M & 2.26 & 2.44 \\
 & FW(1/10)M & 2.56 & 2.81 \\
\hline
10~$\mu$s & FWHM & 1.51 & 1.59 \\
 & FW(1/5)M & 2.31 & 2.41 \\
 & FW(1/10)M & 2.74 & 2.90 \\
\hline
\end{tabular}
\end{center}
\end{table}
Similar results were obtained also with the digital DAQ system.
\subsection{Volume scanning and dead layer thickness}
\begin{figure}[tbp]
\centering
\includegraphics[width=12cm]{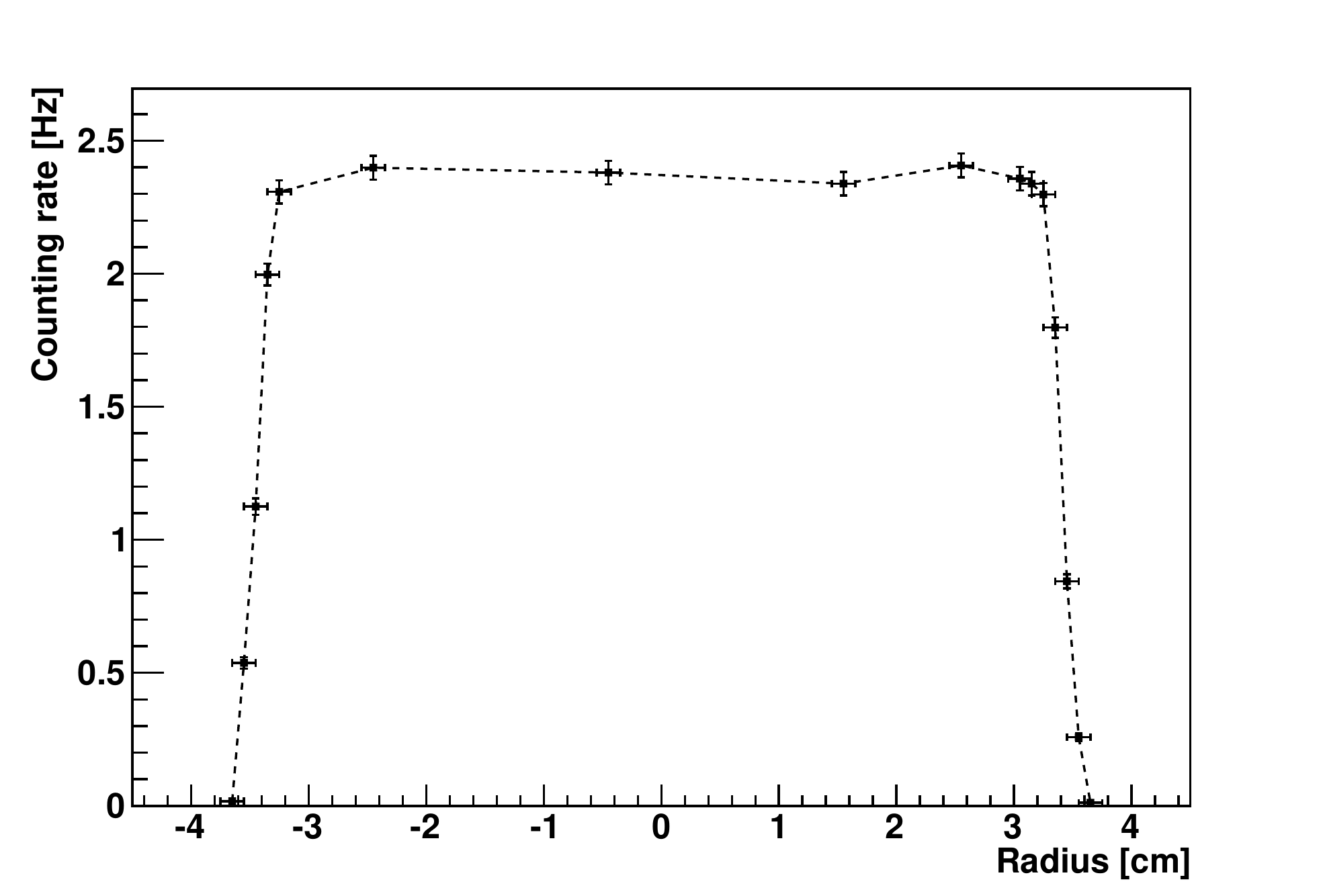}  
\caption{Counting rate at the 60-keV line as a function of radial position for a
collimated $^{241}$Am source placed on the top face of the detector.}
\label{fig:Amscanning}
\end{figure}

\begin{figure}[tbp]
\centering
\includegraphics[width=12cm]{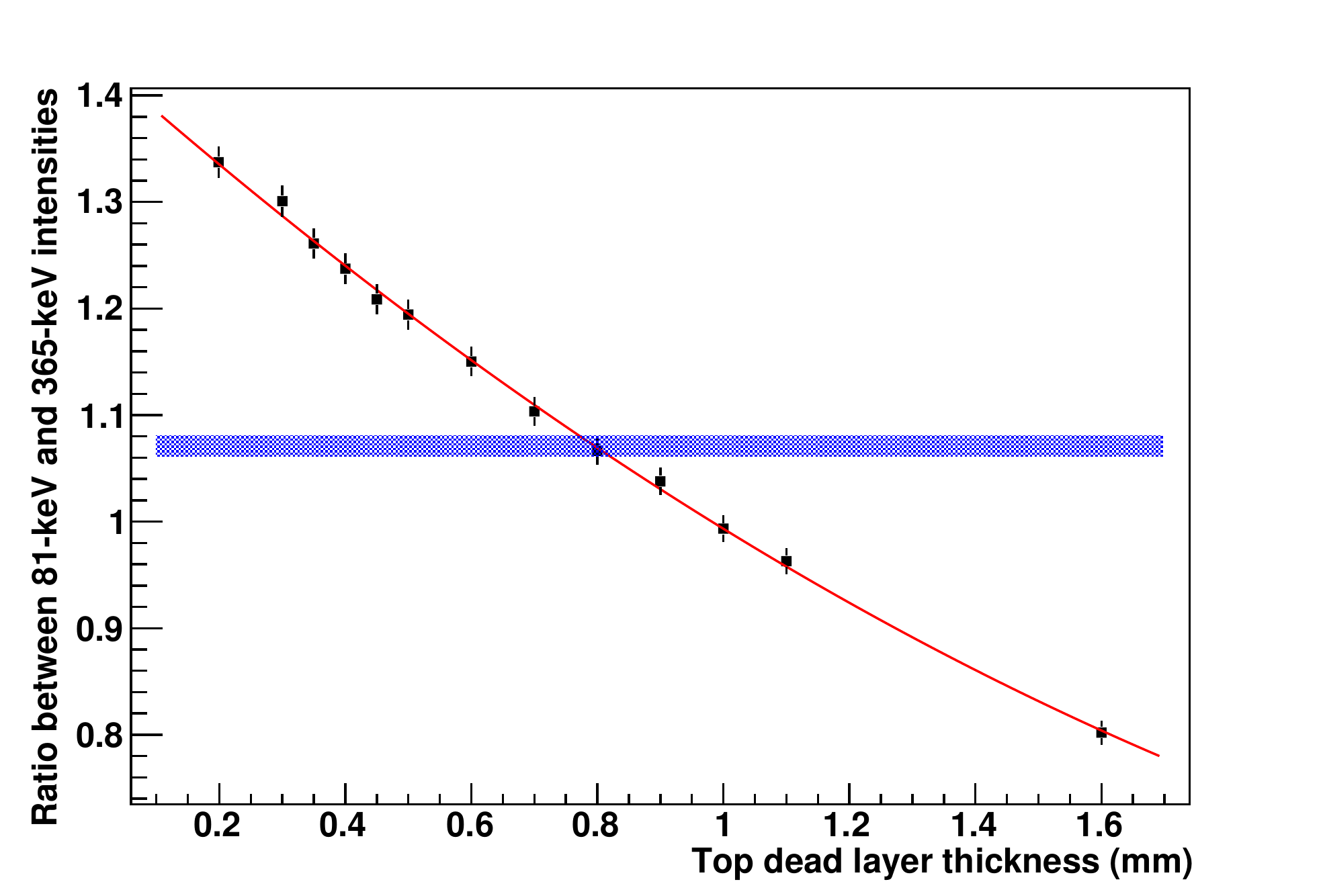}
\caption{Ratio $R$ of the intensities of the 81~keV and 356~keV $\gamma$ lines of
a $^{133}$Ba source as a function of the top surface dead layer thickness, as
evaluated from the Monte Carlo. Error bars account for the statistical 
uncertainty and for the systematic uncertainty due to the description of the detector 
geometry. The values from Monte Carlo have been fitted with a quadratic
curve. The blue shaded area represents the experimental value of the ratio with
1-$\sigma$ uncertainty.}
\label{fig:deadlayer}
\end{figure}

The study of the active volume and of the dead layer features was divided into
two parts. Firstly, the detector active volume and the uniformity of the dead layer 
were investigated by scanning the detector surface with a collimated
$^{241}$Am source.  Secondly, the average thickness of the dead layer was estimated
by irradiating the detector with an uncollimated $^{133}$Ba source.\\
To perform the first set of measurements a mechanical device was built to allow
the positioning of collimated sources with sufficient accuracy along the diameter
of the front face and the side of the detector. The collimator had a hole
of 1~mm diameter and a length of 34~mm.  Using a collimated beam of low-energy
photons (59.5~keV) from the $^{241}$Am source allows to obtain 
well-localized interactions close to the surface.  Consequently, the dead layer
thickness is related to the counting rate in the 59.5~keV line.
Fig.~\ref{fig:Amscanning} shows the counting rate vs. the radial position
on the top surface of the detector. The diameter of the active volume, defined 
as the width with count rate above 50\%(90\%) of the central plateau, 
is 68.6(65.9)~mm. 
The flatness of the central plateau indicates also that the
thickness of the top dead layer is fairly uniform.  
\\
The vertical scanning of the detector yielded less accurate results 
because of the presence of the lateral copper holder and of its reinforcing rings 
(see Fig.~\ref{fig:detgeometry}).
The estimated vertical dimension of the active volume is about 30~mm. 
The detector has been also scanned moving the source along 
a circle centered on the detector symmetry axis, on the top surface 
of the end-cap. Data were used for the validation of the BEGe detector 
modeling and of the pulse shape simulation~\cite{ago10}.\\
The average thickness of the dead layer has been measured by studying the ratio
of the intensities of the $^{133}$Ba $\gamma$-lines at 81~keV and 356~keV. An 
uncollimated 125~kBq $^{133}$Ba source has been used for this purpose.
The ratio $R$ between the two intensities vs. the thickness of the dead layer has
been predicted by means of a Monte Carlo simulation based on the GERDA software
framework MaGe~\cite{mage}. The dead layer thickness on the top of the detector
which is necessary to reproduce the experimental ratio $R= 1.07 \pm 0.01$, as
displayed in Fig.~\ref{fig:deadlayer}, is: 
\begin{equation}
0.79 \pm 0.03 \textrm{\ (stat)} \pm 0.09 \textrm{\ (syst) \ mm}, 
\end{equation}
in good agreement with the manufacturer specifications. The systematic 
uncertainties are due to the Monte Carlo modeling of the detector and to 
the parametrization of the $R$ vs. thickness curve, as shown in 
Fig.~\ref{fig:deadlayer}.
The average dead layer thickness on the side of the detector is 
$0.7 \pm 0.3$~mm (statistical and systematic uncertainties combined in 
quadrature). 
\section{Detector response at different bias high voltages} \label{section:three}
\begin{figure}[ptb]
\begin{center}
\includegraphics[width=12cm]{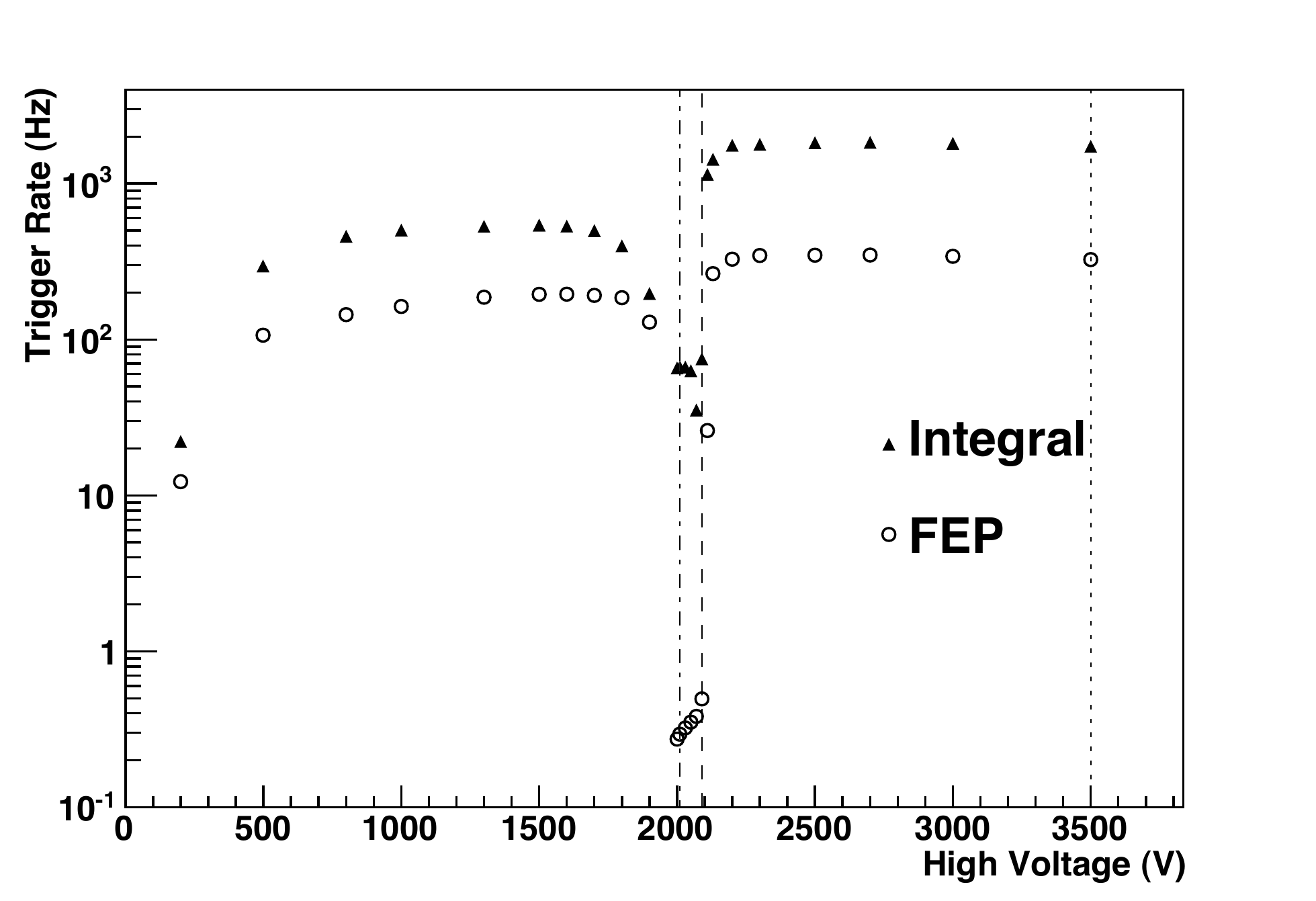} 
\caption{Integral count rate and count rate in the $^{137}$Cs full energy peak
(FEP) at different bias voltage values. Both of them drop drastically in a small
region around 2 kV. The vertical dashed lines mark the values 2010, 2090 
and 3500~V, which are discussed in detail in the text.}
\label{fig:rate}
\end{center}
\end{figure}

\begin{figure}[ptb]
\begin{center}
\includegraphics[width=16cm]{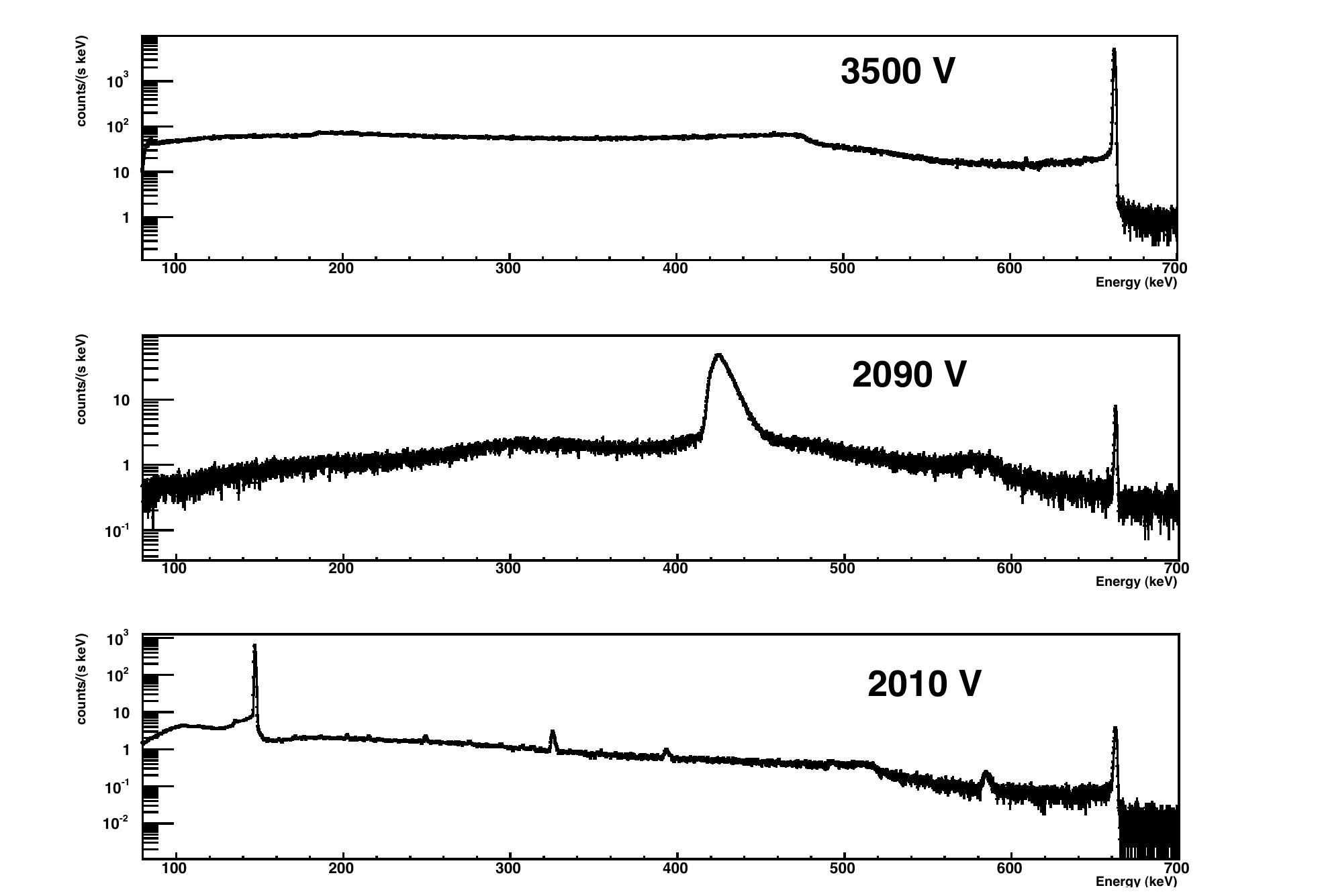}  
\caption{Energy spectrum 
of a $^{137}$Cs $\gamma$ source in different high voltage configurations 
of the detector, namely 3500~V (recommended bias voltage), 2090~V and 2010~V. 
Pulses have been acquired by the digital electronics chain and amplitude is 
reconstructed off-line.}
\label{fig:spectra} 
\end{center}
\end{figure}

The BEGe detector response as a function of the bias high 
voltage (HV) was studied by irradiating the Ge crystal with an uncollimated
$^{137}$Cs source, which emits a single $\gamma$-ray line at 662.3~keV. 
The source was placed on the front face of the detector, on its vertical 
symmetry axis, a few mm above the end cap.
Since the source is uncollimated and the total attenuation length of  
662-keV $\gamma$-rays in metallic germanium is about 27~mm, the detector 
volume illumination is approximately uniform.
%
Charge pulses have been collected with the digital DAQ system 
described in Sect.~\ref{sec:elec}. For this particular dataset, 
pulses have been sampled for 100~$\mu$s 
(trigger after 50~$\mu$s, for a solid baseline estimation) at 
100~MHz rate.
Energy is hence reconstructed from the charge pulses 
according to the Jordanov algorithm~\cite{jo94}, using an integration time of
12~$\mu$s. 
\\
The amplitude for the $^{137}$Cs full-energy peak (FEP) and the counting rate 
were measured different for bias high voltages. Starting from the nominal operation 
voltage of 3500~V  the HV was decreased. The total count rate and the rate in the 
FEP remain
fairly constant down to 2300~V, as shown in Fig.~\ref{fig:rate}. This 
suggests that the actual depletion voltage of the detector is about 2300~V, 
a value much lower than the manufacturer specification.
\\
When the HV is decreased below 2300~V, only a modest and smooth degradation is 
observed, as expected for HV values below the depletion voltage.
But the further reduction of the HV below 2100~V causes a sudden change in the detector operation.
In a voltage interval of about 100~V the integral counting rate is reduced by a factor of $\sim$20
with respect to what is observed above 2200~V. Similarly, the counting rate at
the FEP is suppressed by three orders of magnitude with
respect to the rate at 2200~V.
Below this anomalous voltage interval, the counting rate returns close to the value
before the dip and resumes the previous smoothly decreasing trend. 
Such a peculiar behavior has been reported in previous works~\cite{div09,bu09c} but 
- at our knowledge - never discussed nor experimentally investigated in detail. 
In addition, it has been verified that other BEGe detectors supplied by the same
manufacturer differing only for the dimensions do not show such a peculiar behavior
in their HV curves~\cite{piv10}.\\
Fig.~\ref{fig:spectra} shows the energy spectra (normalized for the live 
counting time) generated by the $^{137}$Cs source at three different bias
voltages: 3500~V (nominal), 2090~V and 2010~V. In all cases, pulse amplitude 
is converted to energy using the calibration data acquired at 3500~V. Energy 
calibration at 2010 and 2090~V should be hence regarded as indicative only.
\\
It can be seen that 
for HV values in the anomalous interval, the FEP at 662~keV is strongly suppressed
in rate, while its position is practically unaffected.
The suppression factor is 700 at 2090~V and 1100 at 2010~V.
In addition, new spurious peaks appear at lower energies, with much higher
counting rate than the FEP. 
At 2090~V there is a broad peak at 426~keV equivalent energy, whose FWHM is about 12~keV, 
while at 2010~V (and in general, in the range 2000-2070~V) a narrow peak appears at
$147$~keV equivalent energy\footnote{Notice, that other small spurious peaks visible at 2010~V 
are due to $\gamma$ lines at higher energy from environmental radioactivity. For 
instance, the line at about 325~keV equivalent energy is related to 
the 1460~keV $\gamma$-ray from $^{40}$K, which is present in the background spectrum.}. \\

\begin{figure}[ptb]
\begin{center}
\includegraphics[width=12cm]{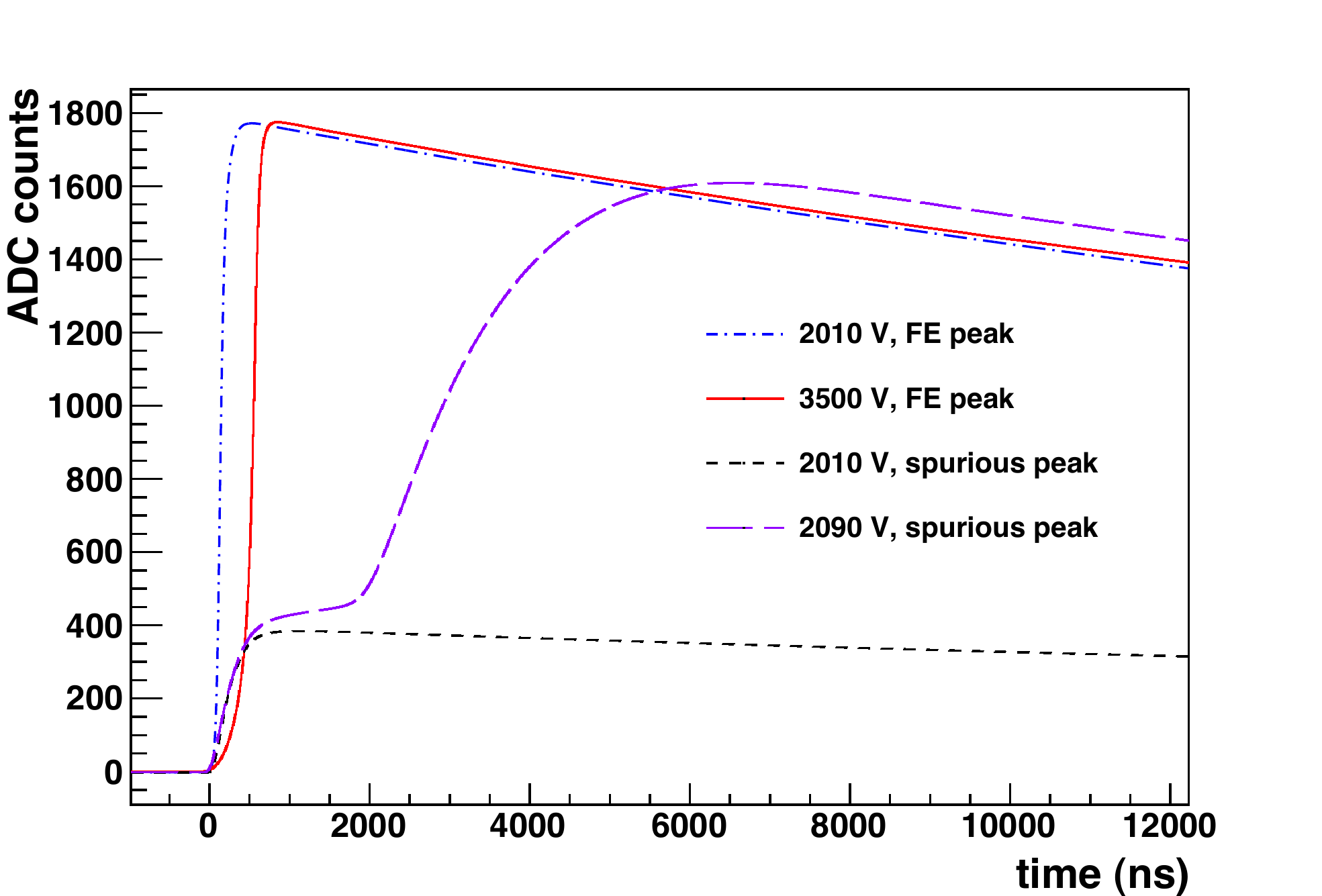} 
\caption{Average pulses from 
events belonging to the structures of the spectra in Fig.~8, 
namely: FE peaks at 2010~V and 3500~V; spurious peaks at 2010~V and 2090~V.}
\label{fig:pulses} 
\end{center}
\end{figure}

To investigate the origin of these new structures, we studied the shapes of the
signals populating the different peaks.
Fig.~\ref{fig:pulses} shows the average signals accumulated for the spurious
and for the full energy peaks. The averaging of the signals provides a pulse shape
which is statistically representative of the peak population and allows to reduce the noise.
The average pulse accumulated for the FEP at 3500~V is an example of a common pulse
expected from BEGe detectors. The signal has a slow rising part at the beginning
followed by a leading edge at the end~\cite{ago10}. The average 10\%-90\% 
rise time of the normal pulses  is about 300~ns.\\
The pulses populating the 426~keV spurious peak at 2090~V are characterized by 
two rising parts connected by a plateau between 0.4 and 2~$\mu$s. 
While the first rising part (0--400~ns) is fast, the second one (2--4~$\mu$s) 
is extremely slow. 
For such a long collection time, part of the charge is expected to be lost by
trapping. 
However, the  pulse has a total amplitude comparable with
the signals of the normal events populating the FEP at 3500~V.
Besides the charge loss by trapping, the underestimation of the peak energy and its 
asymmetric shape is due to
ballistic deficit since the shaping parameters are tuned for faster pulses. 
The broadening may be due either to the analysis algorithm
or - more likely - to high-voltage instabilities. In this regime small
fluctuations of the bias voltage have a significant impact on the peak
position. \\
Differently, the spurious peak at 2010~V originates from 
fast signals (rise time of about 400~ns) which have an amplitude 
clearly smaller (by a factor of five) than the signals creating the FEP. 
The shape of the pulses of the FEP at 2010~V is similar to that observed 
at 3500~V, even if the average rise time is significantly faster 
(rise time is about 200~ns).\\
%
%
%

This peculiar behavior was investigated by using the simulation described in
Ref.~\cite{ago10} and was connected to the electric field
configuration inside the detector.
As shown in Ref.~\cite{ago10}, the electric field of BEGe detectors
operated at the nominal bias voltage (e.g. 3500~V) reaches 
its minimum strength in the middle of the detector.
The holes generated by interactions close to the $n^{+}$ electrode are collected in
the center of the detector and then drift along the same trajectory to the small
read-out electrode (``type I'' trajectories in Ref.~\cite{ago10}). Therefore, the
last part of the signals is independent of the interaction position for most of the
detector volume ($\sim 93\%$ of the total active volume).
The interactions occurring close to the small $p^+$ electrode originate a second
class of pulses (``type II-III'' trajectories in Ref.~\cite{ago10}). These
signals have a faster rise time because the holes generated are not collected in
the center of the detector but directly at the read-out electrode.\\
Decreasing the HV (starting from the operational bias voltage) results in a
reduction of the electric field strength inside the detector. 
When the electric field approaches zero the recombination
probability of the holes increases~\cite{sm} leading to a change in the charge
distribution inside the Ge crystal which contributes to the total electric
field.
The starting of this recombination process ($\sim2300$~V for our detector)
defines the depletion voltage of the detector.
Even if our simulation tools can not provide an accurate computation of the
electric field in the anomalous voltage interval, we estimate that 
the electric field in the middle of the detector is extremely weak for HV values
between 2100~V and 2300~V (``configuration A'').
Moreover, we qualitatively expect that around 2000~V the electric potential has two minima
placed in the small electrode surface and in the center of the detector, respectively.
In this configuration both the small electrode and the center of the detector act as collection
sites for the holes (``configuration B'').
Below 2000~V the two collection sites merge creating a conducting non-depleted
region which expands when the HV is further decreased. 
This region approaches the $p-n^+$  junction when the bias voltage is removed.\\
This model can explain the features of the average pulses in Fig.~\ref{fig:pulses}.
In the ``configuration A'' the holes generated by ``type I'' events are
collected in the center of the detector and then drifted to the read-out
electrode.
The plateau of the pulses in the spurious peak at 2090~V corresponds to the slowing
down of the hole drift in the middle of the detector where the electric
field is weak.
In the ``configuration B'', part of the holes are collected and trapped 
in the middle of the detector and do not reach the read-out electrode. 
The corresponding signals have a smaller amplitude, defined by the
value of the weighting potential in the middle of the detector~\cite{ago10}. 
The reduction of the signal amplitude observed in Fig.~\ref{fig:pulses} 
is in good agreement with the simulation results.
Moreover, the amplitude of this signal corresponds to the amplitude at the plateau 
of the pulses in the spurious peak at 2090~V. 
In fact, the second collection site in the middle of the detector should
originate in the region of low electric field at 2090~V.
\\
According to our model, only the ``type I'' trajectories which pass through the
center of the detector should populate the spurious peaks. 
The pulse shape for events
close to the read-out electrode (``type II-III'') is only slightly affected by 
differences in the electric field due to different bias voltages; therefore, 
these events eventually populate the FEP.
This is consistent with the fact that the FEP is always present - at the same 
position - also for HV values in the anomalous interval, although its rate is 
suppressed, as displayed in Fig.~\ref{fig:spectra}. Furthermore, the rise time 
of the FEP pulses at 2010~V is shorter than that at 3500~V because only 
the ``type II-III'' signals are present.\\

To estimate the volume of the detector originating the spurious peaks we
performed a set of measurements with a collimated $^{137}$Cs source, scanning
the top surface and the side of the detector. 
The spurious peaks are strongly suppressed when only the region
close to the small read-out electrode is irradiated. They are instead enhanced
when the top surface of the detector is illuminated.
We qualitatively estimated that the volume originating the FEP at the
anomalous HV values has a radius and a height of the order of 1~cm. 
The region is a few percent of the total active volume and corresponds to the
region which is expected to originate ``type II-III'' pulses~\cite{ago10}.\\
The peculiar behavior of the considered BEGe detector (dip in the
HV response, pulse shapes) is due to a specific combination of geometrical
and intrinsic characteristics (e.g. the impurity concentration) and it is not
observed in some other BEGe detectors provided by the same manufacturer. 
For instance, the dip in the HV curve is not observed in two BEGe detectors
produced by Canberra with Ge isotopically depleted in $^{76}$Ge~\cite{bu10} and
characterized within the GERDA activities~\cite{piv10}; the detectors are modified versions of
Canberra BE5030/S, having 74.5~mm active diameter and 33~mm height. 
\section{Pulse shape analysis} \label{section:four}
\begin{figure}[tbp]
\centering
\includegraphics[width=12cm]{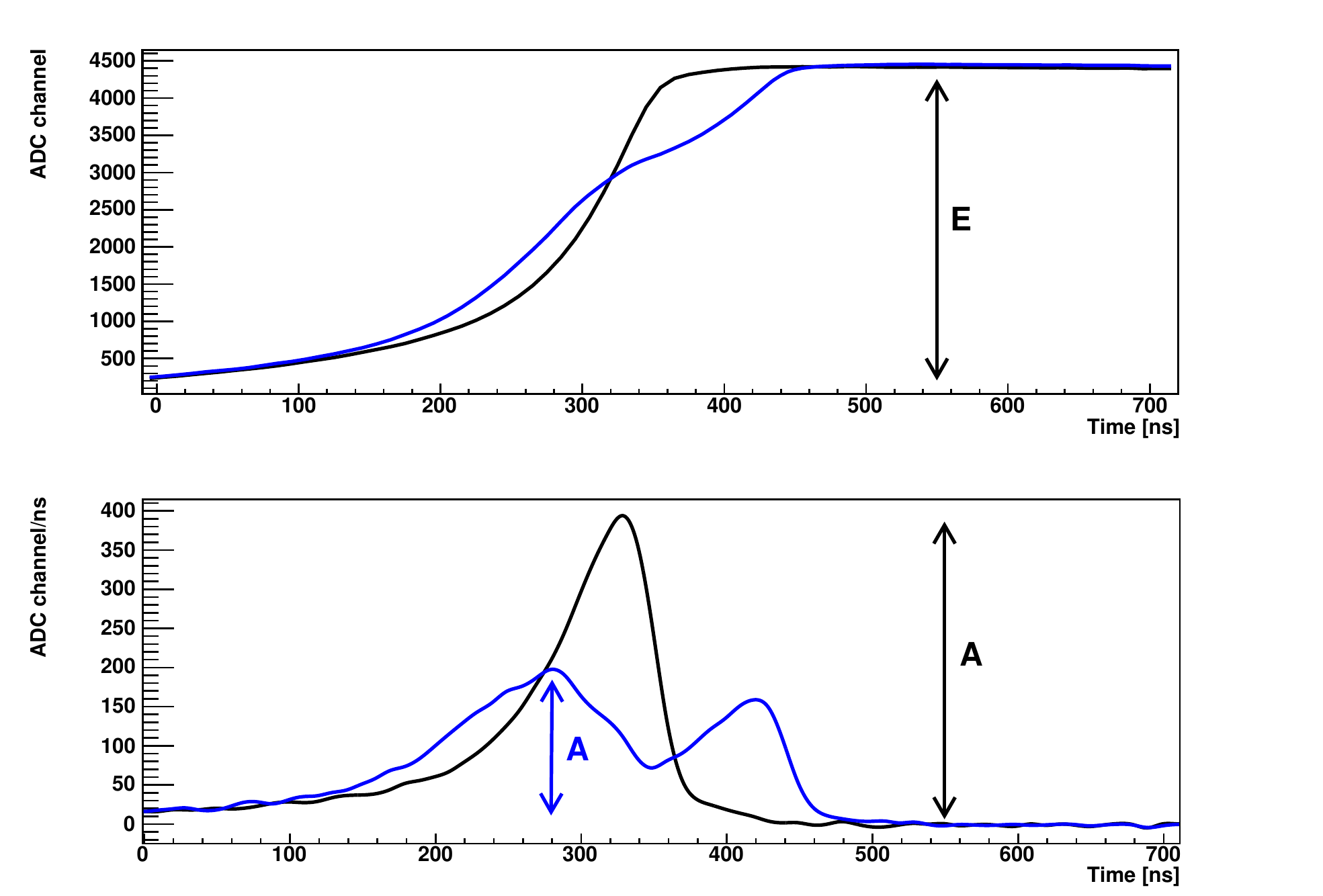}
\caption{Comparison of the signal generated by a multi-site (blue lines) and by a
single-site event (black lines) of
approximately the same energy (about 2300~keV). Upper and lower panels show the
charge and current pulses, respectively.  The current pulse is obtained by numerical
differentiation of the charge pulse. The amplitudes $E$ of the two pulses as
calculated by the off-line reconstruction correspond to 2317.4~keV (SSE, black)
and 2335.1~keV (MSE, blue), respectively.  It is clearly visible that the charge
pulse from a MSE is a superposition of two smaller pulses and that multi-site
and single-site current pulses have very different maxima, in spite of the
similar energy.}
\label{SSEvsMSE}
\end{figure}

\begin{figure}[!htb]
\centering
\includegraphics[width=12cm]{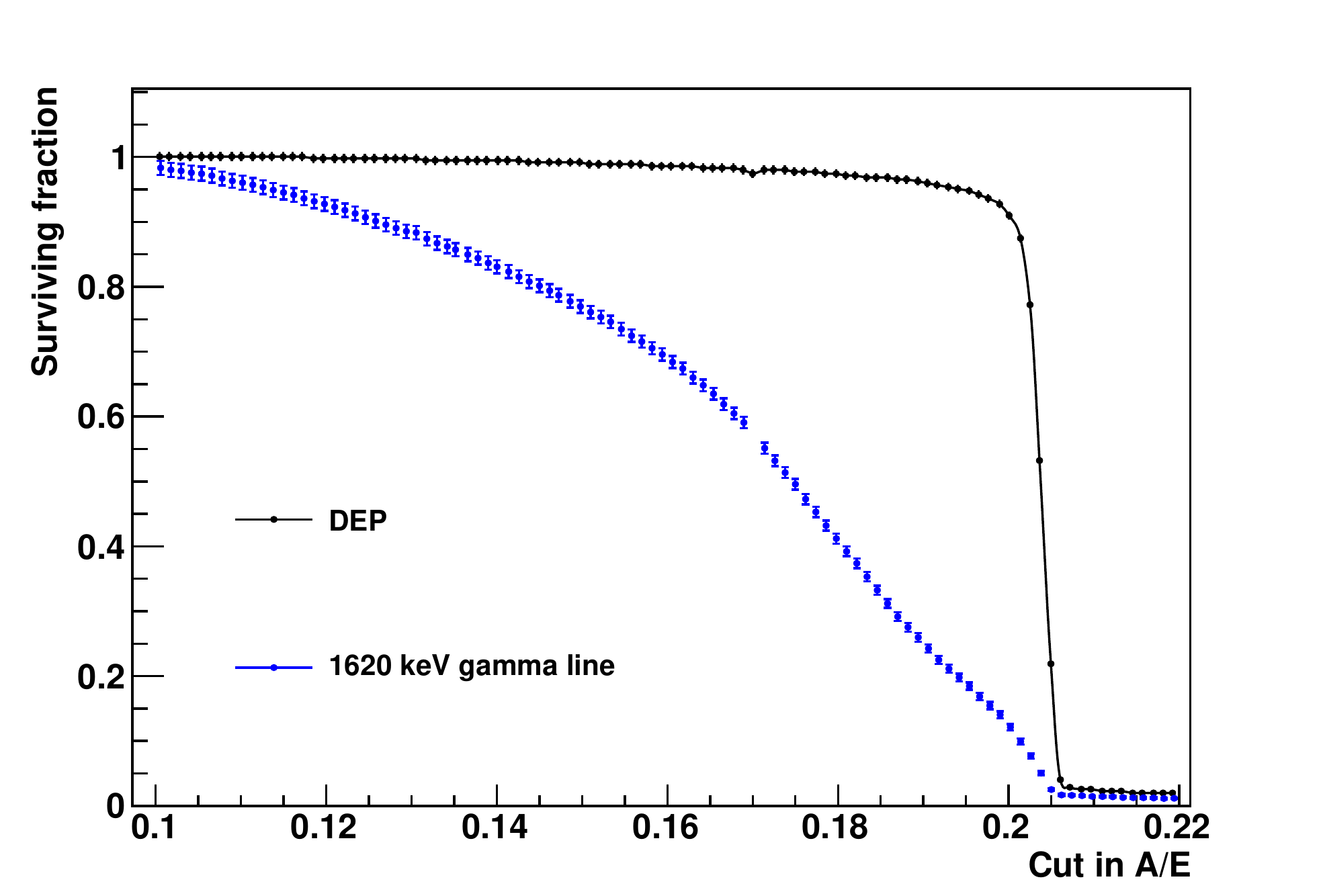}  
\caption{Fraction of events surviving the $A/E$ pulse shape discrimination vs.
the threshold $(A/E)_{thr}$. Analysis is done for the DEP (black, upper line) of
the 2614-keV line and for the $^{212}$Bi $\gamma$ line at 1620~keV (blue, lower
line). The Compton continuum has been statistically subtracted.}
\label{surviving}
\end{figure}

\begin{figure}[htb]
\centering
\includegraphics[width=12cm]{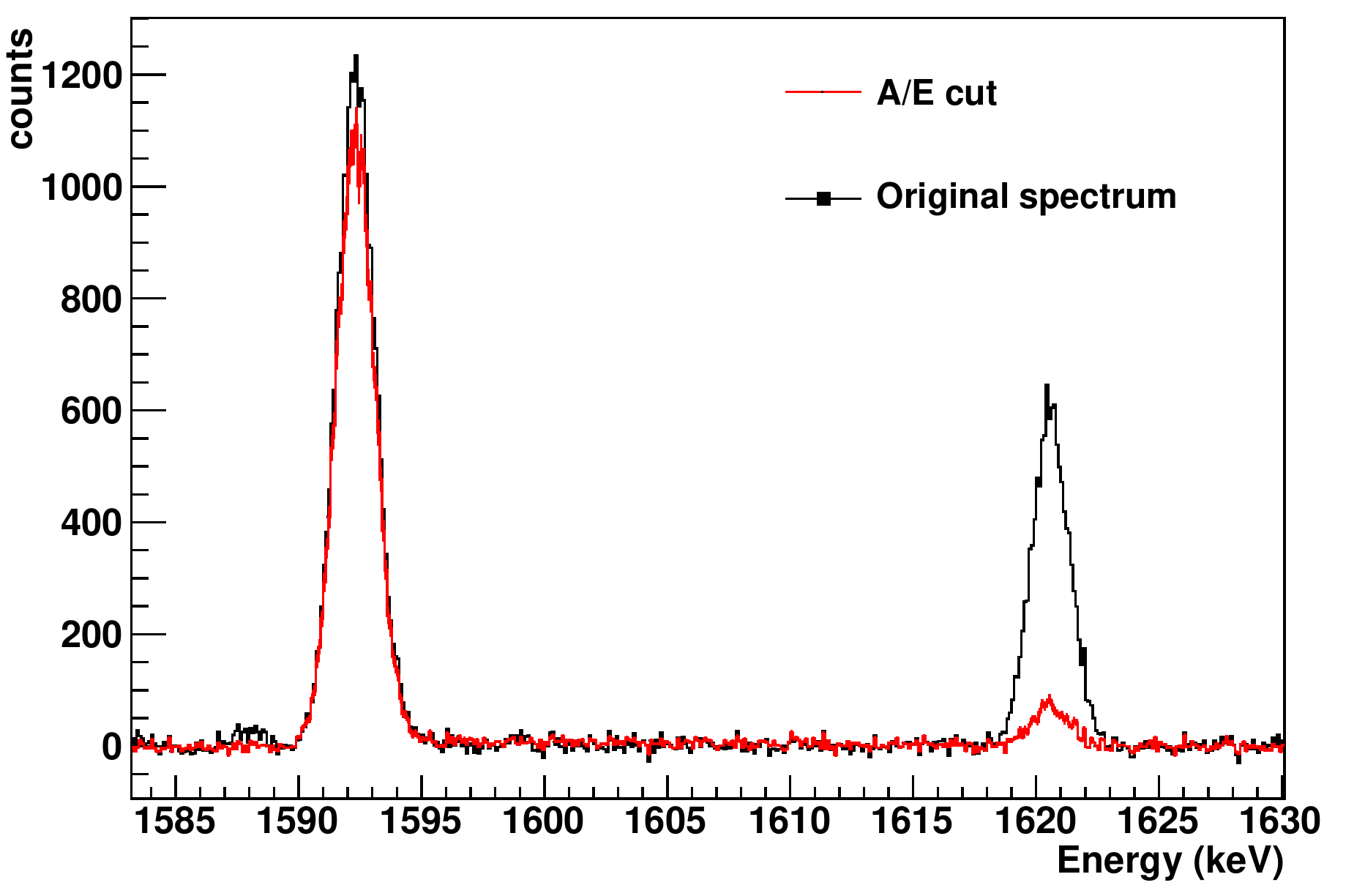}  \caption{Comparison of intensities of 
the DEP 
and of the $\gamma$ line from $^{212}$Bi of the $^{228}$Th spectrum before and 
after the cut based on 
the $A/E$ parameter (black and red spectrum, respectively). 
In both cases the continuum Compton background has been 
subtracted. The threshold on $A/E$ has been chosen to obtain a 90\% survival 
fraction in the DEP.} \label{cut}
\end{figure}

The excellent performance of BEGe detectors in discriminating single-site
events (SSE) from multi-site events (MSE) was recently reported~\cite{bu09,bu09b}.
A SSE is characterized by a single energy deposition localized in a
$\lesssim1~\mbox{mm}^3$ volume. On the other hand, MSE consist of several
interaction sites separated by a typical distance of the order of 1~cm
(e.g. Compton scattering).\\
Such a feature is of primary interest for experiments like GERDA looking for neutrinoless
double beta decay (DBD). Indeed, genuine DBD events are generated by two
electrons that have a range of less than 1~mm in germanium and hence belong
to the former category of events. The background due to
$\gamma$-rays is typically multi-site, because $\gamma$-rays in the energy range
of interest mainly undergo Compton scattering with mean free path 
of a few centimeters in germanium. 
Therefore, an efficient MSE vs. SSE discrimination allows to reject
$\gamma$-induced background, while preserving genuine DBD events.\\
In Ref.~\cite{bu09,bu09b} a discrimination method based on the ratio
between the maximal current $A$ (maximal amplitude of the current pulse)
and the total energy $E$ released in the crystal was proposed and validated. 
As shown in Fig.~\ref{SSEvsMSE}, the events can be classified as single-site if the 
value of the parameter $A/E$ is higher than a given threshold $(A/E)_{thr}$.  More details
about this subject are provided in Ref.~\cite{ago10}.
\\
The rejection power of our detector has been tested irradiating the Ge crystal
with a $^{228}$Th source. The maximum of the current pulse $A$
has been calculated numerically from the digitized charge pulses, after having
applied a 50-ns average filter to reduce the noise. 
The fraction of surviving events as a function of the
value of the threshold $(A/E)_{thr}$ has been determined for the double escape peak (DEP) of
the 2614-keV  $\gamma$-line of $^{208}$Tl  and for the
full-energy  $\gamma$-ray peak at energy 1620.5~keV from the $^{212}$Bi decay. 
In fact, the sample in the DEP is highly enriched in single-site events, 
since it is originated by
the full absorption of the kinetic energy of the e$^+$-e$^-$ pair created by
the $\gamma$-ray when both annihilation quanta escape. Therefore, the peak energy is 
E$_{DEP}$=$E_{\gamma}-2m_{e}c^{2}=1592.5~$keV. Being impossible to 
distinguish double escape from Compton scattering events in the same energy range 
on an event-by-event basis, the populations are separated statistically, by 
fitting the energy spectrum with a Gaussian peak plus a linear background. 
Similarly, the $^{212}$Bi peak is dominated by multi-site events and is used 
to estimate the rejection efficiency of $\gamma$-ray events. In the 
following survival probabilities are quoted with respect to the peaks only, 
namely having subtracted the Compton continuum. 
As shown in Fig.~\ref{surviving}, $(A/E)_{thr}$ can be tuned to achieve an
acceptance as high as 90\% for the DEP, while keeping only 10\% of the 
$^{212}$Bi peak.
The comparison of the DEP line and the $^{212}$Bi line before and after the
$A$/$E$ cut when the DEP acceptance is fixed at 90\% is shown in Fig~\ref{cut}.
\\
Keeping the DEP acceptance fixed at 90\% the discrimination power has been
evaluated  also in other energy regions of interest, i.e. the $^{208}$Tl full
energy peak (FEP) at 2614~keV, the single escape peak (SEP) of $^{208}$Tl
(E$_{SEP}$=2103 keV) and the Compton continuum in a 20-keV region 
around the $Q_{\beta\beta}$ value 
of $^{76}$Ge (2039~keV). 
\begin{table}[tp]
   \caption{Surviving fraction of events from different populations of 
interest after pulse shape discrimination 
   based on the $A/E$ parameter. Acceptance of the DEP is set to 90\%. \label{PSD}}
   \begin{center}
   \begin{tabular}{| c | c | c | c |}
   \hline
   FEP & SEP & FEP & Q$_{\beta\beta}$ \\
   2614 keV & 2103 keV & 1620 keV & 2039$\pm$10 keV \\
   $\lbrack \%\rbrack$ & $\lbrack \%\rbrack$ & $\lbrack \%\rbrack$ & $\lbrack \%\rbrack$ \\
   \hline
   6.4$\pm$0.1 & 6.2$\pm$0.4 & 11.5$\pm$0.1 & 37.5$\pm$0.5\\
   \hline
   \end{tabular}
   \end{center}
\end{table}
The results are summarized in Tab.~\ref{PSD}: 
the intensity of full-energy $\gamma$-ray lines can be
suppressed by a factor of 10, while the Compton continuum can be reduced by a
factor of 3 in the region of interest around the $Q_{\beta\beta}$ value of
$^{76}$Ge. \\
The continuum at $Q_{\beta\beta}$ is in general a mixture of SSE and MSE, 
produced by single- and multiple-Compton scattering of 
higher energy $\gamma$-rays, respectively. The 
survival probability after the $A/E$ cut 
for events in the continuum at $Q_{\beta\beta}$ 
is hence related to the fraction of genuine 
single-site Compton events. On the other hand, the ratio of the SSE and MSE 
populations in the Compton continuum depends on the specific configuration 
of the $^{228}$Th source and of the experimental set-up, including the external 
shielding. The acceptance of the $A/E$ cut for genuine single-site events at 
$Q_{\beta\beta}$ could not be assessed experimentally, but it can be estimated 
by Monte Carlo simulations, as reported in Ref.~\cite{ago10}.
%
\section{Conclusions} \label{section:five}
The performance of a 630~g commercial BEGe detector (model B3830/S by Canberra)
has been systematically investigated and found to be excellent. A peculiar
behavior has been observed in its response as a function of the bias high
voltage: for biases around 2000~V (to be compared to the nominal depletion
voltage, 3000~V) there is a narrow range of about 100~V where charge collection
is strongly reduced.  This is related to the electric field
configuration inside the crystal, since in that range of bias voltages a very
weak field is expected to be present in the central part of the detector, that
extends up to collecting electrode and prevents the full charge collection. An
interesting feature of BEGe detectors, mainly for experiments looking for
neutrinoless double beta decay, is their pulse shape discrimination capability
of single-site vs. multi-site events based on the $A/E$ parameter. It is possible 
to reject more than
90\% of $\gamma$-ray events while having 90\% of surviving events for genuine
SSE.  These results are in agreement within a few percent with the ones obtained
in Ref.~\cite{bu09} with a detector of the same type, but with a larger diameter
(80 mm). Note also that spectroscopic performance is remarkably stable in a
wide range of bias high voltage above 2400~V.

\acknowledgments
We would like to thank our GERDA Collaborators for very useful discussions, as
well as the staff of the Gran Sasso laboratory for the precious help and
support.


\begin{thebibliography}{99}
\bibitem {bege}
CANBERRA Broad Energy Ge (BEGe) Detector, catalog accessed at URL:
\href{http://www.canberra.com/products/485.asp}{http://www.canberra.com/products/485.asp}

\bibitem {canberra}
Canberra Semiconductor NV, Lammerdries 25, B--2430 Olen, Belgium.

\bibitem{div09} A.~di Vacri et al., \emph{Characterization of Broad Energy Germanium 
Detector (BEGe) as a candidate for the GERDA experiment}, 
\emph{IEEE Nucl. Sci. Symp., Conf. Record 2009} (2009), 1761.

\bibitem{ago09} M.~Agostini, \emph{Characterization of a Broad Energy Germanium detector 
through advanced pulse shape analysis techniques for the GERDA double-beta 
decay experiment}, Master Thesis, Universit\`a di Padova (2009).

\bibitem{bu09}
D.~Budj\'a\v{s}, M.~Barnab\'e Heider, O.~Chkvorets, N.~Khanbekov and 
S.~Sch\"onert, \emph{Pulse shape discrimination studies with a Broad-Energy Germanium 
detector 
for signal identification and background suppression in the GERDA double beta decay experiment}, 
\emph{JINST} \textbf{4} (2009), P10007.

\bibitem{bu09b}
D.~Budj\'a\v{s}, M.~Heisel, W.~Maneschg and H.~Simgen, 
\emph{Optimisation of the MC-model of a p-type Ge-spectrometer for the 
purpose of efficiency determination}, {\emph{Appl.\ Radiat.\ Isot.} {\bf 67} (2009) 706}.

\bibitem{ba10} M.~Barnab\'e Heider, D.~Budj\'a\v{s}, K.~Gusev  and S.~Sch\"onert,  
\emph{Operation and performance of a bare broad-energy germanium detector in liquid argon}, 
\emph{JINST} \textbf{5} (2010), P10007.

\bibitem{ge04}
GERDA Collaboration, I.~Abt et al.,
\emph{GERDA:
The GERmanium Detector Array for the search of neutrinoless
$\beta \beta$ decay of $^{76}$Ge at LNGS}, Proposal,
\href{http://www.mpi-hd.mpg.de/ge76}{http://www.mpi-hd.mpg.de/ge76}.

\bibitem{ago10}
M.~Agostini et al., \emph{Signal modeling of HP-Ge detectors with a small
read-out electrode and application to neutrinoless double beta decay search in 
$^{76}$Ge}, companion paper, submitted to \emph{JINST}, 
preprint [arxiv:1012.4300].

\bibitem{bu09c}
D.~Budj\'a\v{s}, \emph{Germanium detector studies in the framework of the GERDA experiment}, 
Ph.~D. thesis, University of Heidelberg (2009). 

\bibitem{co10}
R.~Cooper and D.~Radford, talk given at 
``Workshop on Germanium-Based Detectors and Technologies'', 
Berkeley, May 18-20, 2010.

\bibitem{ortec} ORTEC Advanced Measurement Technology, Inc., catalog at 
\href{http://www.ortec-online.com/Solutions/gamma-spectroscopy.aspx}{http://www.ortec-online.com/Solutions/gamma-spectroscopy.aspx}

\bibitem{caen1728} See URL 
\href{http://www.caen.it/nuclear/product.php?mod=N1728B}{http://www.caen.it/nuclear/product.php?mod=N1728B}.

\bibitem{tuc} See URL 
\href{http://www.iphc.cnrs.fr/-TUC-.html}{http://www.iphc.cnrs.fr/-TUC-.html}.
%

\bibitem {ge11}
M.~Agostini, L.~Pandola, P.~Zavarise and O.~Volynets,
\emph{GELATIO: a general framework for modular digital analysis of HPGe
signals}, in preparation.

\bibitem{mage}
M.~Bauer et al., \emph{MaGe: a Monte Carlo framework for the Gerda and Majorana double 
$\beta$ decay experiments}, {\emph{J.\ Phys.\ Conf.\ Ser.} {\bf 39} (2006) 1}; \\
M.~Boswell et al., \emph{MaGe: a Geant4-based Monte Carlo application framework for 
low background germanium experiments}, submitted to \emph{IEEE, Trans. Nucl. Scie.}, 
preprint [arxiv:1011.3827.v1].

\bibitem{jo94}
V.T.~Jordanov and G.F.~Knoll, \emph{Digital synthesis of pulse shapes in real time for 
high resolution radiation spectroscopy}, {\emph{Nucl.\ Instrum.\ and\ Meth.\ A} 
{\bf 345} (1994) 337}.

\bibitem{piv10} G.~Pivato, \emph{Experimental characterization of a Broad Energy Ge 
detector for the GERDA experiment}, Master Thesis, Universit\`a di Padova (2010). 

\bibitem {sm}
   S.M.~Sze, \emph{Physics of Semiconductor Devices}, John Wiley \& Sons, 1981.\\
   S.~Selberherr, \emph{Analysis and Simulation of Semiconductor
   Devices}, Springer-Verlag, 1984.

\bibitem{bu10} M.~Agostini et al., \emph{Procurement, production and testing of BEGe detectors 
in $^{76}$Ge}, to appear in Nucl. Phys. B, Proc. Suppl. (Neutrino 2010).

\end{thebibliography}
\end{document}